# An application of physical units (dimensional) analysis to the consideration of nonlinearity in electrical switched circuits

*Emanuel Gluskin*

The Kinneret College on the Sea of Galilee (The Jordan Valley), Braude Academic College (Carmiel), Holon Technological Institute, and Ben-Gurion University, Israel.
gluskin@ee.bgu.ac.il,  http://www.ee.bgu.ac.il/~gluskin/

**Abstract**:  The argument of physical dimension/units is applied to electrical switched circuits, making the topic of the nonlinearity of such circuits simpler.  This approach is seen against the background of a more general outlook (IEEE CAS MAG, III, 2009, pp. 56, 58, 60-62) on singular (switched and sampling) systems.  The appendix suggests some points for the axiomatization of system theory considered as an independent science.

## 1.  Introduction

### 1.1.  General

   As is well known, linear and nonlinear systems possess different properties. Switching circuits and systems are no exception to this view [5], and the study of the conditions for their linearity and nonlinearity is interesting and timely because, being suitable technologically, such systems have become a common performance/design tool.

   The "switching nonlinearity" can be obtained without any typical "analytical nonlinearity" overstress of the magnetic or ferroelectric, etc., material.  The elements being switched can be the usual linear time invariant (LTI) elements, specified for voltage and current stresses that are much higher than the stresses actually appearing in the switched unit, even if this unit exhibits a strong nonlinearity in its functioning. That is, there is no problem of reliability of the elements, which is not so for a ferromagnetic-core inductor or a ferroelectric-layer capacitor for which any exhibition of a strong nonlinearity (usually needed for a not-too-weak nonlinearity of the whole circuit) means an overstress as regards the specification of the element.

   The "switching nonlinearity" is obtained just by proper control of the switching, and is exhibited in each time interval in which at least one such properly controlled switching occurs.  Thus, it is technologically no more difficult to obtain nonlinear switched circuits than linear (LTV) switched circuits.  This circumstance positively influences both the applications and the theoretical role of the nonlinear circuits.

### 1.2. The state-space outlook of [5]

Let us first recall the position of [5] which the outlook developed in the present work methodologically completes.

The general situation is [5] that the switching operation leads either to a *nonlinear* (*NL*), or a *linear time-variant*, (*LTV*) circuit. If the instances of switching are defined by state-variables $\mathbf{x} = \{x_k\}$, i.e. by the functions (voltages or currents) that have to be determined, then the system is NL, and if the switching instances are prescribed, i.e. known a priori, then the system is LTV. One readily sees this distinction equationally.

In the first case, the space-state equations for the state vector $\mathbf{x}(t)$ can be presented as

$$d\mathbf{x}/dt = [A(\mathbf{t^*(x)})]\mathbf{x} + [B(\mathbf{t^*(x)})]\mathbf{u}(t) , \qquad (1)$$

where $\mathbf{t^*}$ are the switching instants, here defined by the state variables ($\mathbf{x} \rightarrow \mathbf{t^*}$), and $\mathbf{u}$ are the inputs. More simply written, (1) becomes

$$d\mathbf{x}/dt = [A(\mathbf{x})]\mathbf{x} + [B(\mathbf{x})]\mathbf{u}(t) \qquad (1a)$$

and is obviously nonlinear.

In the second case, each moment of singularity $t^*$ is actually defined by some known time function $f(t)$; $f(t) \rightarrow t^*$, or $t^* = t^*(f(t))$, simply written as $t^*(t)$, and we have

$$d\mathbf{x}/dt = [A(\mathbf{t^*}(t))]\mathbf{x} + [B(\mathbf{t^*}(t))]\mathbf{u}(t) \qquad (2)$$

or

$$d\mathbf{x}/dt = [A(t)]\mathbf{x} + [B(t)]\mathbf{u}(t) \qquad (2a)$$

which is obviously linear.

When considering the very fact that we deal with a switching system, both (1) and (2) can be written as

$$d\mathbf{x}/dt = [A(\mathbf{t^*}(.))]\mathbf{x} + [B(\mathbf{t^*}(.))]\mathbf{u}(t) , \qquad (3)$$

which means that the situation of linearity or nonlinearity depends on the control of the switching instants. It may be said that leaving in (3), i.e. in the notation $\mathbf{t^*}(.)$, the *possibility* of $\mathbf{x} \rightarrow \mathbf{t^*}$ is the essence of the outlook on linearity and nonlinearity of switched systems developed in [5] and references given there.

The point is not so much to obtain an analytical solution for $\mathbf{t^*}(.)$; the numerical values $\mathbf{t^*}$ are some easily electronically "on line"-*measurable* (detectable) functionals of the functions $\mathbf{x}(t)$, and as is explained below, the very map $\mathbf{x} \rightarrow \mathbf{t^*}$, i.e. a dependence of $\mathbf{t^*}$ on $\mathbf{x}$ can be easily stated (created) in the switched system.

In some circuit situations the nonlinear case can be named "feedback control" of the switching instants or of the values of the switched elements, which are included in the matrices in (1) (see [5] for more details), but it is important to see that the components of $\mathbf{x}$ controlling the switching can be *any* state variables of the system, not necessarily its intended outputs.

This classification of dynamic equations with time singularities as linear or nonlinear can be relevant to different fields of science. As a somewhat informal example, being closest to operational research, consider the state (of the blood pressure, or simply the mood, etc.) of a patient in a hospital as a time function $x(t)$, and assume first that giving medicines for all the patients (**x** denotes their states) in the hospital is done at prescribed time instances **t***, i.e. independently of **x**. Then the system (process) is of the type "[A($t$)]", i.e. *linear*, as (2,2a).

Assume now that the medicines are given each time when it is individually requested by the patients. This obviously means **x** → **t***, or "[A(**x**)]" in terms of (1,1a), and the system/process is *nonlinear*. Only in the latter case can a "chaos" (or, rather, some critical complexity that can lead to a dangerous or even a catastrophic disorder in the hospital) occur.

Looking around, one can observe different examples of linearity and nonlinearity through such "singular system thinking", i.e. through analysis of the nature of the points of singularity of the physical processes. Work [5] even explains the tendency to chaotic movement in a statistical ensemble of particles by a dynamic nonlinearity of the "switching type". That is, this type of nonlinearity is clamed to be typical in many natural processes.

### 1.3. How can one create the switching nonlinearity in an electronic circuit?

Work [5] and references there explain that the most suitable method of creating the dependences **t***(**x**) is by using electronic comparators having at least one of their inputs as an $x(t)$. The outputs of the comparators are some pulses triggering the switches at the instants **t***, and thus a circuit realization of the nonlinear map **x** → **t*** is obtained. See Fig. 1.

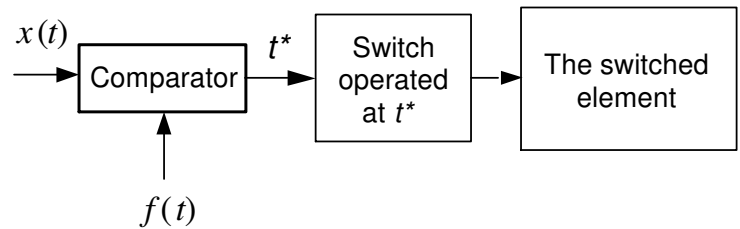

Fig. 1: A switching *subsystem* somewhere inside a system under study. The comparator's input $x(t)$ causes nonlinearity of the whole system, because the switching instant $t^*$ and the switched at $t^*$ element (e.g. a capacitor, and thus also the matrix [A] in (1), which includes the capacitance) depend on $x(t)$ as in (1). The switching instant $t^*$ appears as the instant of the crossing by $x(t)$ the reference function $f(t)$ for which the options ((a)-(c)) are given in the main text. Some switching pulse (not shown in the figure) is generated at $t^*$ to operate the switch.

In such a system, a switch can be operated at the instant when a state variable $x(t)$ (graphically) crosses some level (the other input of the comparator) that can be:

(a) a constant level (most usually),

(b) a known (prescribed) function $f(t)$,

(c) another state variable.

In each of these cases, the instant of switching is defined by $x(t)$, and in case (c) even by two state variables.

We thus create nonlinearity by means of *directly* comparing some $x(t)$ to a fixed level or to another function.

In [14] another way is found; namely some functionals of the state variables are created (*calculated* on line), and the switching instances are defined when these functionals reach some prescribed levels. This is, of course, also a form of the map **x** → **t\***, i.e. the **t\***(**x**)-nonlinearity, though this nonlinearity of the *whole system* in which some linear subsystems are thus switched is not observed in the analysis in [14]. Work [6] makes this basic point clear, making it possible to refer to the system of [14] as to a nonlinear system, and seeing the chaotic state obtained there as a result of the switching nonlinearity, i.e. just in view of (1,1a).

It is interesting to know how to create the **t\***(**x**)-nonlinearity in different ways, and many other switching circuits and systems, e.g. those also appearing in [2-4] and references cited there, should be considered in terms of the "**t\***(.)-outlook".

The method of dimensional analysis of the "switching nonlinearity" to which we now pass on cannot compete with the described **t\***(.)-outlook in its generality, but completes it by introducing some methodological simplicity.

## 2. The nonlinearity of a switched circuit in terms of dimensional argument

Let us speak about an "analytically compact" circuit in which all state variables are mutually connected. For such a circuit any switching causes a singularity in any of the state variables (though these singularities may be very different), and we can speak about any $x(t)$.

Let us consider, for such a circuit, the operationally simplest nonlinear case of a state variable $x(t)$ crossing a constant (critical) level $x_{cr}$, i.e. item (a) in Section 1.3. This crossing defines the instance $t^*$ at which a switch has to be operated (Fig.1), as a solution of the equation

$$x(t^*) \ = \ x_{cr} \ . \tag{4}$$

Thus found, $t^*$ obviously depends on the parameters defining $x(t)$, which is the sense of the map **x** → **t\***.

The physical dimension of $x_{cr}$ is that of $x(t)$, and if $x(t)$ contains an amplitude-type parameter $x_0$ (acquired from an initial condition, or an input function, -- below we give examples for both cases) of the same dimension, then we have to consider the non-dimensional ratio $\dfrac{x_o}{x_{cr}}$ and the inevitable, in view of (4), dependence of $t^*$ on this ratio.

While for a linear system, for which no component of **x** is involved in any switching control, it must simply be

$$x(t) \sim x_\mathrm{o} \ ,$$

when $x_{cr}$ is involved, the dependence of $x(t)$ on $x_\mathrm{o}$ (and thus the whole system under study) becomes nonlinear. Indeed, considering the physical dimensions, we have to write

$$x(t) \ = \ x_o \, F(\frac{x_o}{x_{cr}}; \ \lambda_1, \lambda_2, ...; \ \frac{t}{\tau_1}, \frac{t}{\tau_2}, ...) \ , \tag{5}$$

having included in the appearing function $F$: the ratio $x_o / x_{cr}$ that interests us, some non-dimensional constant parameters $\lambda_1, \lambda_2, ...$ characterizing a circuit's structure in some relative or proportional units, and the circuit's time-constants $\tau_1, \tau_2, ... .. .$ In general, $F$ includes all the influencing non-dimensional parameters.

Considering that

$$\tau_2 = \xi_1 \tau_1; \ \ \tau_3 = \xi_2 \tau_1, ...$$

where

$$\xi_1 = \frac{\tau_2}{\tau_1}, \ \xi_2 = \frac{\tau_3}{\tau_1}, ...$$

are non-dimensional parameters of the type $\lambda_1, \lambda_2, ...$, we can include $\xi_1, \xi_2, ...$ into $\lambda_1, \lambda_2, ...$, rewriting (5) as:

$$x(t) = x_o \, F(\frac{x_o}{x_{cr}}; \lambda_1, \lambda_2, ...; \frac{t}{\tau_1}) \ , \quad\quad\quad (6)$$

having only one non-dimensional variable containing time. One may find such inclusion of the time variable more physical.

When wishing to check (6), one can use it in (4), obtaining

$$F(\frac{x_o}{x_{cr}}; \lambda_1, \lambda_2, ...; \frac{t^*}{\tau_1}) = \frac{x_{cr}}{x_o} \ , \quad\quad\quad (4a)$$

which obviously yields (requires) $t^* = t^*(x_o / x_{cr})$, in agreement with the initial point of the argument.

Mutual connection of the time- (here, $t^*$) and the amplitude- (here, $x_o$) type parameters is typical for nonlinear systems, and since many oscillators exhibit nonlinearity via the amplitude-time dependencies in their dynamics, it can be expected that the dimensional argument can be applied, in particular, to the theory of oscillators.

The amplitude-type parameter $x_o$ plays a very important role below. First of all, it is obvious from (6) that if $x_{cr}$ is (as we generally assume) an *independent* parameter, then any influence of $x_{cr}$ in (6) means a nonlinear dependence of $x(t)$ on the input (initial) parameter $x_o$, that is, $x(t)$ and the whole process, or system, are nonlinear.

Furthermore, (6) also simply shows *linearity* of a system, and not only in the trivial cases of no $x_{cr}$ existing, or for $x_{cr} \to \infty$ (i.e. $x_{cr}$ becoming so large that no level-crossing and resultant switching can occur), but also in the nontrivial case when $x_{cr} \sim x_o$ (i.e. $x_{cr}$ is *not* independent) when, in (6), $F$ ceases to be dependent on $x_o$. The importance of the latter case will be illustrated later.

In general, the dimensional argument can classify the systems as linear or nonlinear very simply, but some nontrivial points associated with a commonly used circuit over-idealization can arise, as will be discussed in Section 3.

### 2.1. The first circuit example

The circuit shown in Fig. 2 appears in [5]. Switch $S_1$ is closed at $t = t_1 = 0$, and then $S_2$ is closed at some moment $t_2$.

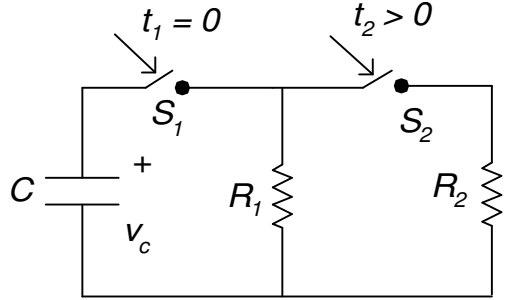

Fig. 2: A circuit considered for two modes of its operation. Switch $S_1$ is closed at $t = 0$, and then $S_2$ is closed *either* at some (any) prescribed/independent moment, *or* when $v_c$ reaches a certain level $v_{cr}$. In the first case, $t_2$ is independent of $v_C$, and the circuit is linear (LTV). In the second case, $t_2$ depends on $v_{cr}$ and thus inevitably on $v_C$ (which, in the terms of (1-4) is our '$x$'), and the circuit is nonlinear.

Based on the "**t\***(.)-argument", we can expect that:

1. If $t_2$ is prescribed, then the circuit is linear, and $v_c(t)$ is directly proportional to $v_o$ over the whole time axis.

2. If the switching of $S_2$ occurs when $v_C(t)$ decreases to a certain level $v_{cr} < v_o$, i.e. *when the instant of switching $t_2$ depends on $v_C$*, then the circuit is *nonlinear*, which is simply seen from the fact that for $t > t_2$, $v_C(t)$ is *not* directly proportional to $v_o$.

Indeed, by direct analysis of the circuit, work [5] shows that in the second case

$$t_2 = \tau_1 \ln \frac{v_o}{v_{cr}} \tag{7}$$

and

$$v_C(t) = v_{cr} \left( \frac{v_o}{v_{cr}} \right)^{\frac{\tau_1}{\tau_2}} e^{-\frac{t}{\tau_2}} ,$$

$$\sim v_o^{\frac{\tau_1}{\tau_2}} e^{-\frac{t}{\tau_2}} , \quad t > t_2 . \tag{8}$$

where $\tau_1 = R_1 C$, and $\tau_2 = R_1 R_2 (R_1 + R_2)^{-1} C < \tau_1$.

Since $\tau_1 \neq \tau_2$, the dependence of $v_C(t)$ on $v_o$, given by (8), is nonlinear for any time interval including $t_2$ (for instance $0 < t < \infty$), and thus it is concluded that for the second case/mode the circuit is nonlinear.

### 2.2. *Application of the dimensional argument to the circuit of Fig.2*

Let us now, not using the precise solution, just by making (6) and the associated arguments less abstract, explain why when $t_2$ is prescribed the above circuit is linear, and in the opposite case – nonlinear.

When $t_2$ is prescribed, i.e. no $v_{cr}$ is introduced; among the given parameters only $v_o$ gives the unit of volt that is necessary for the resulting $v_C(t)$. Thus, by the dimensional reasons, it *must be* that

$$v_C(t) = v_o F(\frac{t}{\tau_1}, \frac{t}{\tau_2}), \quad \forall t, \qquad (9)$$

with non-dimensional function $F(.)$, or, using, in the spirit of (6), the fact that $\frac{t}{\tau_2} = \frac{\tau_1}{\tau_2} \cdot \frac{t}{\tau_1}$, and somewhat changing the structure of '$F$', we have

$$v_C(t) = v_o F(\frac{\tau_1}{\tau_2}, \frac{t}{\tau_1}) . \qquad (9a)$$

This case is linear since $v_C(t) \sim v_o$ .

In the case when $v_{cr}$ is introduced, having both $v_o$ and $v_{cr}$ measured in volts, we have one more non-dimensional parameter, $\frac{v_o}{v_{cr}}$. When this parameter is not too small, i.e. when the equation $v_c(t) = v_{cr}$ has a solution $t^*$, i.e. the switching actually occurs (otherwise existence of the parameter $v_{cr}$ cannot be verified), it influences the process, and we have to modify (9a) to

$$v_C(t) = v_o F(\frac{v_o}{v_{cr}}; \frac{\tau_1}{\tau_2}, \frac{t}{\tau_1}) . \qquad (10)$$

According to (10), the (any) influence of $v_{cr}$ makes $v_C(t)$ nonlinear by $v_o$, and this case is nonlinear. The dimensional argument immediately makes the nonlinearity obvious.

We turn now to a popular, but not quite simple circuit. First, the idealizations leading to $\delta$-functions and thus hiding the realistic possibility of obtaining a nonlinearity, will be considered. Then the case of linearity, mentioned regarding (6), i.e. given by means of the proportion $x_{cr} \sim x_o$, will be illustrated.

### 3.  The dimensional argument and circuits' over-idealizations

In some idealized cases, when $\delta$-type current spikes, caused by sources of finite voltages, are allowed to occur (think about $x(t)$ in view of (4)), it is impossible to directly use comparators of the time functions for creating $\mathbf{t^*(x)}$. The resistors are necessary since we need a *finite slope* of the function $x(t)$ at the level-crossing with $x_{cr}$, for the dependence of $t*$ on the parameters included in $x(t)$, i.e. $\mathbf{x} \rightarrow \mathbf{t*}$, be determined from (4).

In a special form, this problem is also seen in the above circuit analysis, namely, if we let both $R_1$ and $R_2$ be zero in the nonlinear by $v_o$ factor included in (8),

$$v_o^{\frac{\tau_1}{\tau_2}} = v_o^{(1+\frac{R_1}{R_2})} \; , \qquad (11)$$

having $R_1/R_2 \geq 0$ *undefined*, perhaps even zero, then the very fact of nonlinearity becomes vague.

### 3.1. The second circuit

The idea of taking zero resistances in the previous circuit is, however, not practical from any point of view, but if we turn to the circuit in Fig. 3, then we have to face its common interpretation as an SCC modeling a resistor, *just for $R_1$ and $R_2$ zero*.

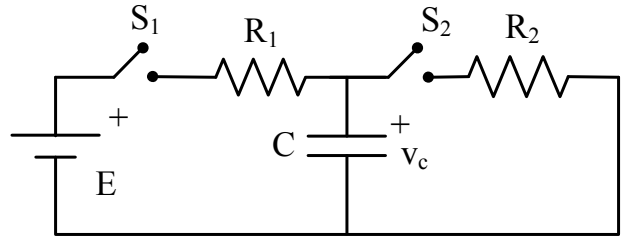

**Fig 3**:  A circuit with a dc source.  $S_1$ and $S_2$ act alternatively, $(\uparrow,\downarrow)$ and $(\downarrow,\uparrow)$, with the same frequency, providing charging and discharging of the capacitor.  Usually, one sets $R_1$ and $R_2$ zero (which is unrealistic for several reasons, even formally *prohibited* [5] *for lumped circuits* by basic circuit theory, concerned with the frequency ranges of the signals, but accepted for the analysis of some applications of this well-known SCC), obtaining $\delta$-type current spikes, making it impossible, in view of (4), to use any voltage levels to influence the moments of starting the charging and/or discharging of the capacitor.  The *over-idealization* of zero $R_1$ and $R_2$, not allowing us to introduce any $v_{cr}$ and a nonlinearity according to (4) and (6), cannot be accepted here.

In the usual version of the circuit, when the resistors are assumed be zero and all the processes are infinitely quick, the switches are operated with the prescribed frequency $f$ which can be "any", and in terms of the average current taken from the battery, one is able to obtain model of the averaged *linear* frequency-controllable resistor $R(f) = (fC)^{-1}$.  The switching instances are ("externally") prescribed here.

This usual case is an idealized one, and there is no place for introducing any $v_{cr}$, in this case.  However, being interested in setting our conditions for linearity and nonlinearity in this switching realization of the resistor, we have to use some $v_{cr}$ for the operation of at least one of the switches.

Thus we assume -- for both the linear and nonlinear versions thus appearing, -- some small realistic nonzero resistors $R_1$ and $R_2$, and using some $v_{cr}$, consider different cases of operation of $S_1$ and $S_2$, which give linear and nonlinear versions of the changed circuit.

This time let us use only the dimensional argument, starting from the nonlinear case.

### 3.2. The dimension analysis of the circuit of Fig. 3

For this circuit, $E$ plays the role of $x_o$ in (6), and according to (6) if the moment of closing of, e.g., $S_2$ depends on some critical level $v_{cr}$, then $v_c(t)$ must depend on $E$ nonlinearly,

$$v_c(t) = E \cdot F(\frac{E}{v_{cr}}, \frac{t}{\tau_1}, ....) . \qquad (12)$$

This nonlinearity means, in particular, that the *equivalent resistor*, "seen" by the battery, is nonlinear.

The introduced resistors strongly change the conditions for the circuit operation. There is no infinitely quick time process now, but we have the usual exponential processes that continue, in principle, for an infinite time, if full charging or full discharging of the capacitor is required. For obtaining a periodic process, we must limit the charging and discharging to certain voltage levels, which are reached by $v_c(t)$. This can be associated with some operational requirements as follows, which can, as well, make *the linear case* not quite trivial, but the dimensional argument demonstrates its effectiveness in this case too.

Let us choose the criterion for the switching of $S_1$, which starts (in a periodic process) the charging, when $v_c(t) = k_1 E$, and the criterion for switching $S_2$, which starts the discharging, when $v_c(t) = k_2 E$. For instance, taking $k_1 = 0.1$ and $k_2 = 0.9$, we have good use of the battery range/voltage *for any E*, which is a practical point. In order to simply see that in the case when each $v_{cr}$ is *only thus*, and not independently, involved, we have a *linear* circuit we substitute the chosen levels of $v_{cr}(E)$, $k_1 E$ and $k_2 E$, into (12) rewritten for this case as

$$v_c(t) = E \cdot F(\frac{E}{v_{cr1}}, \frac{E}{v_{cr2}}, \frac{t}{\tau_1}, ....) . \qquad (12a)$$

The substitution of $v_{cr1} = k_1 E$ and $v_{cr2} = k_2 E$ eliminates $E$ in $F(.)$:

$$v_c(t) = E \cdot F(\frac{1}{k_1}, \frac{1}{k_2}, \frac{t}{\tau_1}, ....) \sim E , \qquad (13)$$

making it obvious that for such operation of the switches the system is linear.

Of course, the same conclusion is obtained if we first rewrite (12) not as (12a), but as

$$v_c(t) = E \cdot F(\frac{E}{v_{cr1}}, \frac{t}{\tau_1}, \frac{v_{cr1}}{v_{cr2}}, ...). \qquad (12c)$$

When using such $v_{cr1}$ and $v_{cr2}$ *in addition* to the independent $v_{cr}$ appearing in (12), we have a nonlinear circuit.

Thus, equation (6) appears to be useful in both proving nonlinearity and linearity in the different possible cases.

## 4. Conclusions

We have compared two new outlooks, a state-space one [5] and the new dimensional one, on the nonlinearity of switched systems. Hopefully, these outlooks can make the classification of systems as linear and nonlinear easier and more physically interesting for many readers, while a system theorist may find some research aspects here. Works [7-11] may be recommended for further reading regarding the **t***(.)-outlook, including (see, especially, [9,11]) the topic of *sampling* circuits. The problem of determination and classification of different possible ways of obtaining the "switching nonlinearity" **t***(**x**) seems to be important for a theoretician and a designer.

Equation (6) allows one to simply see the nonlinearity, but the sometimes present "over-idealization" of circuits should be avoided. The systematic use of scaling factors (as $v_o$ and $E$) in the dimensional analysis is emphasized.

Sources [13,15,16] present a general introduction to switched circuits, and the classical pedagogical work [1] should not be missed.

It is worth noting that the use of dimensional argument is better known in physics than in circuit-theory literature, and for instance, in the many-volume "*Course of Theoretical Physics*" by L.D. Landau and E.M. Lifshitz many interesting examples can be found. However, the traditional use of the dimension argument reveals the qualitative dependence of a solution on all the given parameters, while here, focusing on the scaling factors, we specifically orient this argument to revealing the nonlinearity (or linearity) of a system, to increase the understanding of switched systems. As far as we know, this approach is novel.

Informally, the basic motivation for the research presented in [5-11] and here includes the wish to start to speak about system theory in heuristically useful, simple terms having general educational importance. According to this motivation, the Appendix focuses on some conceptual and logical problems whose proposed discussion should be helpful for the heuristic axiomatization. Hopefully, the constructive line of thought of the Appendix might be accepted by the supporters of a more formal approach, because it is just mathematics that gives us the inspiring example of a theory in the foundation of which an axiomatization is found.

### Appendix:  Some material for a heuristic axiomatization of system theory

This Appendix considers some problematic points related to the very foundation of system theory.  Our opinion is that system theory cannot rely on only some purely mathematical system of axioms, because the *importance* of the linear, or any other mathematical space *cannot be postulated*; it is an empirical fact.  For instance, the linear superposition of forces, acting on the same body, is an empirical fact, and even inside mathematics per se, the role of the axiomatic concepts and schemes was gradually revealed via many "empirical" (analytical and logical) investigations and the development of concrete theories.  Historically, the practice has often led mathematical thought, and system theory can also thus contribute.

#### A.1. Nonlinearity as a basic, independent concept

Sometimes a "nonlinear system" is *defined* as "*not a linear one*".  This seemingly (philologically) perfect definition is logically wrong, because one cannot define anything via something not given.  In the context of our equations, this definition possesses the unacceptable form of the definition of [A(**x**)] as "not [A($t$)]", and in the terms of the characteristics of circuit elements, it becomes the not more attractive definition of a curve as not a straight line.

The use of the expression "not a linear one" arises from the historical role of the relatively simple linear systems that defined the language of electrical engineering.  However, the technological situation has changed; it is similarly easy today to create linear and nonlinear switched and sampling systems, and it is unclear why, when starting to think about a nonlinear effect, we have to recall (by using the words "*not a linear*") linearity at all with its *irrelevant ad hoc* features.

We thus suggest that the term "nonlinear system" be replaced by the terms "**x**-*system*", or "**u**-*system*" (Section A.3), or "**x**-**u**-*system*", according to the actual system situation.  "*Nonlinear* **x**-*system*", etc., is also acceptable; the point just is that a definition has to be constructive.

#### A.2. System structure and dynamic equations

The equational form that we use in the main text,

$$d\mathbf{x}/dt = [A(\mathbf{x})]\mathbf{x} + [B(\mathbf{x})]\mathbf{u}(t) , \qquad (A1)$$

is preferred to the more usual (and even somewhat more general) normal-form

$$d\mathbf{x}/dt = F(\mathbf{x}, \mathbf{u}(t)) , \qquad (A2)$$

because of some inherent connection of the concept of switching with the concept of *structure*.  This connection is generally (phenomenalogically) seen in (A1), but not in (A2).

Indeed, one can come to (A1) ([5] for more details) starting from an LTI system without any switching, and then introducing the switching which changes the elements of this system at **t**\*, thus obtaining [A(**x**)] as [A(**t**\*(**x**))].  This derivation of (A1) preserves the matrices that in the LTI case directly (even if incompletely) reflect the system's structure.  One can thus understand system's nonlinearity as a dependence of the parameters of the system's elements (or of it's structure) on the state variables, i.e. on the physical processes occurring in the system. The heuristically useful (see examples in [10]) structural interpretation of an equation, or a system, and its nonlinearity is less natural for (A2).

The structural point of view should be attractive for a designer to whom the form (A1) can give ideas for *structural generalizations* relevant to modern multi-element *nonlinear* systems. See also [12].

### A.3. System inputs and the nonlinear "**u**-systems"

One agrees that for a mathematical formulation of a system, its "inputs", -- a part of its structure, -- have to be given. However, one should not act under the assumption that this always is a simple point. The following example shows how easily a nonlinear system can be wrongly classified as a linear one, because of ignorance the physical role of a given function.

Consider the scalar equation

$$dx/dt \ + \ a(f(t)) \ x(\mathrm{t}) \ = 0 \qquad\qquad (A3)$$

in which $f(t)$ (as well as $a(.)$) is a *known* function. Let us assume that $f(t)$ is the input (or one of the inputs) of the system that is associated with (A3). This assumption is *unusual*, because just the *right-hand side* of an equation is usually considered as the input function given directly, or via some known operator. However we cannot rely on the nonmathematical concepts of right and left.

If $f(t)$ is the input, then if this system is a linear one, the linear scaling test

$$(f(t) \to k\,f(t)) \ \Rightarrow \ (x(t) \to k\,x(t)), \qquad\qquad (A4)$$

where $k$ is a constant, has to succeed. However, if $a(.)$ is not constant, this test obviously does not pass in (A3). (Note that it is possible, and sometimes more suitable, to see (A4) thus: the *simultaneous substitutions* $f \to kf$ and $x \to kx$ do not change the equation.) Thus, if $f(t)$ is the (an) input, (A3) is a nonlinear equation.

Assume now that the known function $f(t)$ is not any input, i.e. it does not belong to a *set* (a linear space over the numerical field) of some known functions from which we freely pick up $f(t)$, or $kf(t)$, or some $k_1 f_1(t) + k_2 f_2(t)$, etc., but is a *fixed* function, inherent for such a system/device. Then, we can assume, as is usually done, that *the right-hand side of (A3) is the input function that, in this particular case, is identically zero*, i.e. we have the usual LTV equation of the type

$$dx/dt \ + \ a(f(t)) \ x(t) \ = u(t), \qquad\qquad (A5)$$

where $u(t)$, and not $f(t)$, is the input. (An inclusion of $u(t)$ into $a(.)$ would create nonlinearity as in (A3).)

This argument is easily generalized to vector equations. The equation

$$d\mathbf{x}/dt \ = [A(\mathbf{u})]\mathbf{x} + [B]\mathbf{u}(t) \qquad\qquad (A5a)$$

is *nonlinear* (and not LTV), despite the fact that the input-vector $\mathbf{u}(t)$ may include only known functions. (Again, the freedom in the choice of $\mathbf{u}(t)$ is important.) This is the "**u**-nonlinearity", or the "**u**-system".

The rule here is that an equational term of the type

$$x(t)f(t) ,\qquad\qquad (A6)$$

in which $x(t)$ is the unknown to be found, and $f(t)$ is a known function, *is linear if $f(t)$ is not an input, and is nonlinear if $f(t)$ is an input.*

Thus, without careful analysis of the system's macroscopic structure, which reveals what are the 'inputs', one can easily make a mistake in considering the linearity or nonlinearity of a system or equation.

The fact that modern electronics technology creates systems with very complicated multi-element structures that can be seen as some mutually connected subsystems with numerous inputs involved, requires that the logical system description, preceding the analytical investigation, be done very carefully. One sees that the concept of *structure* becomes a basic, perhaps axiomatic, concept of modern system theory.

### A.4. *Engineering realization, and the proposed symbol* $(f_1 + f_2)_r(\cdot)$.

Another concept worth discussing in the axiomatic sense is the engineering concept of *realization*, especially in its connection with mathematical notations. We shall conduct the relevant discussion around the very basic mathematical notation $(f_1 + f_2)(\cdot)$ for $f_1(\cdot) + f_2(\cdot)$. There is no objection here to the wide use of the symbol $(f_1 + f_2)(\cdot)$ that both makes the writing of the sum more compact, and also overviews the whole function as a function of the same argument, by mathematicians. The point is just to legitimize seeing this "whole function" more as if by an empirical scientist or engineer, than by a mathematician. This leads to some necessary stresses, and can encourage an engineer to indeed use, -- which is not natural for him as it stands now, -- the symbol $(f_1 + f_2)(\cdot)$.

For any mathematician, the expression $(f_1 + f_2)(\cdot)$ is *just* a notation for $f_1(\cdot) + f_2(\cdot)$. Still the physical nature of objects is ignored and no other heuristic arguments appear; this symbol indeed cannot be anything but the notation for the sum. However, let us consider the fact that the equality

$$(f_1 + f_2)(\cdot) = f_1(\cdot) + f_2(\cdot) \qquad\qquad (A7)$$

also represents the important *property of linearity of the sampling ("evaluating") operator acting on physical functions*. (In the present context, it is important that we speak about *time* functions.)

Indeed, the δ-function's sampling action

$$f(t*) = \int_{t*-a}^{t*+a} f(t)\delta(t - t*)dt, \quad a > 0 , \qquad\qquad (A8)$$

is the operator of the map

$$T:\ t \rightarrow t* \ (\text{or } f(t) \rightarrow f(t*)) \qquad\qquad (A9)$$

applied to the $f(t)$. Because of the linearity of the integral, taking in it $\alpha f_1(\cdot) + \beta f_2(\cdot)$, with $\alpha = 1$ and $\beta = 1$, (this is sufficient for the argument), we have

$$T(f_1 + f_2) = Tf_1 + Tf_2 , \qquad\qquad (A10)$$

i.e. we obtain, using (A8), that

$$(f_1 + f_2)(t^*) = f_1(t^*) + f_2(t^*) . \qquad (A11)$$

Since we can perform the sampling at each point of the range of the functions' definition (i.e. $t^*$ is as arbitrary as $t$ ), (A7) is the same as (A11).

The context of the sampling operator shows that $(f_1 + f_2)(\cdot)$ is not just a notation or a definition, because any notation (definition) can be, in principle, given *arbitrarily*, and by itself cannot lead to any essential conclusion, in this case that of the linearity of an operator. One should agree that the feature of linearity of a real sampling device, or any other device, is something important.

Using now the physical terms, let us interpret $(f_1 + f_2)(\cdot)$ not as a notation for the purely mathematical form $f_1(\cdot) + f_2(\cdot)$, but as a symbol for the *realized* summation. Then, the left- and the right-hand sides of (A7) obtain different meanings: '+' in $f_1(\cdot) + f_2(\cdot)$ is the "mathematical order", while '+' in $(f_1 + f_2)(\cdot)$ is this "order" having been realized already.

To make the point feasible, let us assume that the linearity of a real electrical circuit is checked for the principle of superposition, using two input voltage generators, generating some known voltages $v_1(t)$ and $v_2(t)$. In order to observe/check superposition of the circuit response, we have to perform *three independent experiments*: one with $v_1(t)$, one with $v_2(t)$, and the third with $v_1(t) + v_2(t)$, as the input functions of the system.

For the third experiment, we *have to build* (*realize*) a generator of the summed voltages, *this generator to be applied to our system*. Such a generator can (actually, *will*, in order to be sure in correctness of the numerical side) be a *series connection* of the given generators of $v_1(t)$ and $v_2(t)$, i.e. we *create, -- by connecting the plus of one generator to the minus of the other (a series connection), -- an electrical circuit*, whose output voltage is $(v_1 + v_2)(t)$ .

However when seeing $(v_1 + v_2)(t)$ just as a notation for $v_1(t) + v_2(t)$, we have not created a circuit.

This physically reasonable understanding of $(f_1 + f_2)(\cdot) = f_1(\cdot) + f_2(\cdot)$ as an equality connecting the characteristics of *three physical objects* appears to be useful in [11] where (A7) (see eq. (2) in [11]) is considered for formulating the distinction between linear and nonlinear *samplings*, which is the constructive point of [11]. Namely, when the sampling instants are dependent on the function being sampled, then (A11), i.e. (A7), ceases to be correct, and for analysis of the (nonlinear) sampling a special point of view is developed.

Thus, both the sampling operator and the hardware points of view suggest understanding of the symbol $(f_1 + f_2)(\cdot)$ not in the sense of some compactly written $f_1(\cdot) + f_2(\cdot)$, but in the sense of *realization* of the sum.

In order to avoid contradiction between the classical and the present interpretations, introducing the special notation

$$(f_1 + f_2)_r(\cdot)$$

where the subscript '$r$' means "realized", is suggested here.

Of course, the *calculated* sum is some $(f_1 + f_2)_r(\cdot)$ too. In particular, since the result of a *numerical* calculation occupies some *special* locations in the computer's memory, it is a physically independent object, according to our interpretation of (A7).

### A.5. The singular switched and sampling systems and the "real-valued" mathematics

Let us observe that though complex-plane methods are widely used in the theory of switched systems, this theory is more inherently connected with the classical theory of the real variable. The use of electronic comparators (Fig.1) means the use of the mathematical comparisons ">" and "<" (here $x(t) > f(t)$, or $x(t) < f(t)$), which are defined for *real*, and *not* for complex values and functions. (Just see the comparator's action as a realization of Dedekind's "golden cut" A|B that defines the real number via the concepts of ">" and "<"). We detect and directly use only the zeros (or level-crossings, our **t**\*) of *real* time-functions, and the same can be said about the also very important sampling systems ([11] and references there devoted to "Lebesgue's sampling").

It is also important to note that the theory of chaotic systems, with its numerous arithmetical constructions, also basically belongs to the theory of the real variable, and even is historically leading in changing the traditional focus (over more than a century) of the circuit theory on only the methods of complex analysis. From the present general point of view, this circumstance defines the importance of chaos theory even more than its skilful analytical developments do.

### A.6. The pedagogical aspect

The fact that system theory became more "real-valued" should be helpful in developing its popularization. Considering system theory as an independent science, one can follow the example of mathematicians and physicists who have written not a few well-known excellent popular books, making basic mathematics and physics *a part of one's general culture*.

Today, no such parallel is observed in system theory literature. However, Laplace called his famous book "*The System of the World*", and such possible titles of popular books as, for instance, "*System theory for everybody*", "*'Systems' around you*", "*Instabilities and robustness in society*", "*You as your system*", and "*The curve that wants to be beautiful at each of its points*" might motivate system specialists to write popular educational texts. Already the example of the hospital situation, from which we started, can lead one to many other relevant examples. See also [12].